\documentclass[aps,prd,preprintnumbers]{revtex4}
\usepackage{graphicx}
\graphicspath{{./figures/}}
\usepackage{epsfig}

\begin{document}

\preprint{HU-EP-07/43, ADP-07-12/T652}

\title{The gluon propagator from large asymmetric lattices}

\author{O. Oliveira}
\affiliation{CFC, Department of Physics, University of Coimbra, 3004 516 Coimbra, Portugal}
\author{P. J. Silva}
\affiliation{CFC, Department of Physics, University of Coimbra, 3004 516 Coimbra, Portugal}
\author{E.-M. Ilgenfritz}
\affiliation{Humboldt-Universit\"at zu Berlin, Institut f\"ur Physik, Newtonstr.~15, 12489 Berlin, Germany}
\author{A. Sternbeck}
\affiliation{CSSM, School of Chemistry \& Physics, The University of Adelaide, SA 5005, Australia}

\begin{abstract}
          The Landau-gauge gluon propagator is computed for the 
          $SU(3)$ gauge theory on lattices up to a size of $32^3
          \times 200$. We use the standard Wilson action at $\beta =
          6.0$ and compare our results with previous computations
          using large asymmetric and symmetric  lattices. 
          In particular, we focus on the impact of the lattice geometry
          and momentum cuts to achieve compatibility between data from
          symmetric and asymmetric lattices for a large range of
          momenta.
\end{abstract}

\maketitle

\section{Introduction and motivation}

The investigation of the Landau-gauge gluon propagator
\begin{equation}
  D^{ab}_{\mu\nu} (q) ~ = ~ \delta^{ab}
       \left( \delta_{\mu\nu} - \frac{q_\mu q_\nu}{q^2} \right) \, 
       D(q^2)
\end{equation}
in lattice QCD dates back to more than 20 years. 
Despite the efforts made by a number of authors, there are questions which 
remain to be answered. For large momenta, let us say $q > 2$ GeV, the results 
from different groups provide a consistent picture of the propagator and agree
well with calculations performed using other non-perturbative techniques. 
On the other hand, the infrared limit is still an open issue and has been a 
field of intense research in the last years (see the contributions to this
conference, {\it e.g.}~\cite{Latt07Proceedings}, and references therein). 
From the point of view of lattice simulations, the questions
to be answered yet are numerous and sometimes are even not easy to
handle with. For example, it is still under debate how to deal with the
Gribov ambiguity in lattice simulations,
and how close present lattice data come to Landau-gauge gluodynamics
in the continuum and infinite-volume limit.

In order to access the infrared limit of the gluon propagator, two of us 
(O.O. and P.J.S.) recently 
proposed and explored the use of large asymmetric 
lattices \cite{Bra,Arx,Sard}, {\it i.e.} $L^3 \times T$ with $T \gg L$. 
The price of relying on such kind of lattices are the control, or the lack 
of it, of additional finite-volume effects coming from a breaking of the $Z_4$ 
symmetry, a remnant of the $O(4)$ continuum symmetry on a symmetric 
hypercubic lattice. 
When previous studies on symmetric lattices have shown 
strong finite-volume effects in the infrared region \cite{Att}, the situation 
is more dramatic for asymmetric lattices. For example, there the gluon 
dressing function $Z(q^2) = q^2 D(q^2)$ computed at equal time-like and 
spatial momenta are not necessarily compatible within pure statistical errors 
in the low-momentum region.

On the other hand, the access to very low momenta is much more 
computationally intensive in simulations on symmetric lattices compared to 
those on asymmetric ones. Therefore, if somehow the asymmetry-induced 
finite-volume effects were brought under control, data at much lower momenta 
than currently available could be obtained. 
Indeed, then still the infinite volume limit 
has to be taken, but in a situation where more data were available 
in the infrared momentum region.

Having now access to a considerably larger spatial volume for the asymmetric 
case, in this study we report on some first results obtained comparing data on 
symmetric and asymmetric lattices, namely $16^3 \times 256$, $18^3 \times 256$,
$32^3 \times 200$ and $32^4$. In particular, we look for regions 
in the lattice momentum space where
the differences between time-like and spatial momenta disappear and
where not. For our simulations we use the standard Wilson
gauge action with $\beta=6.0$ fixed. This value corresponds to an
inverse lattice spacing of about $a^{-1} = 1.94 {\rm~GeV}$.
To relate our data 
at the different lattice momenta to their continuum counterparts we use 
\begin{equation}
q_{\mu} ~ = ~ \frac{2}{a} \, \sin \left( \frac{\pi n_\mu}{L_\mu} \right) \, ,
 \hspace{1cm} n_\mu = 0, \, 1, \, \dots , L_\mu - 1 ,
\end{equation}
where $L_{\mu}$ is the lattice extent in direction $\mu$. Definitions and
details on the gauge fixing are given in \cite{PRD} and for the $32^3
\times 200$ data in \cite{Stern05}. In the following, whenever
possible, a $Z_3$ average over equivalent momenta is performed.

\section{Asymmetric lattice: $32^3 \times 200$}

\begin{figure*}
\vspace*{0.8cm}
\includegraphics[scale=0.3]{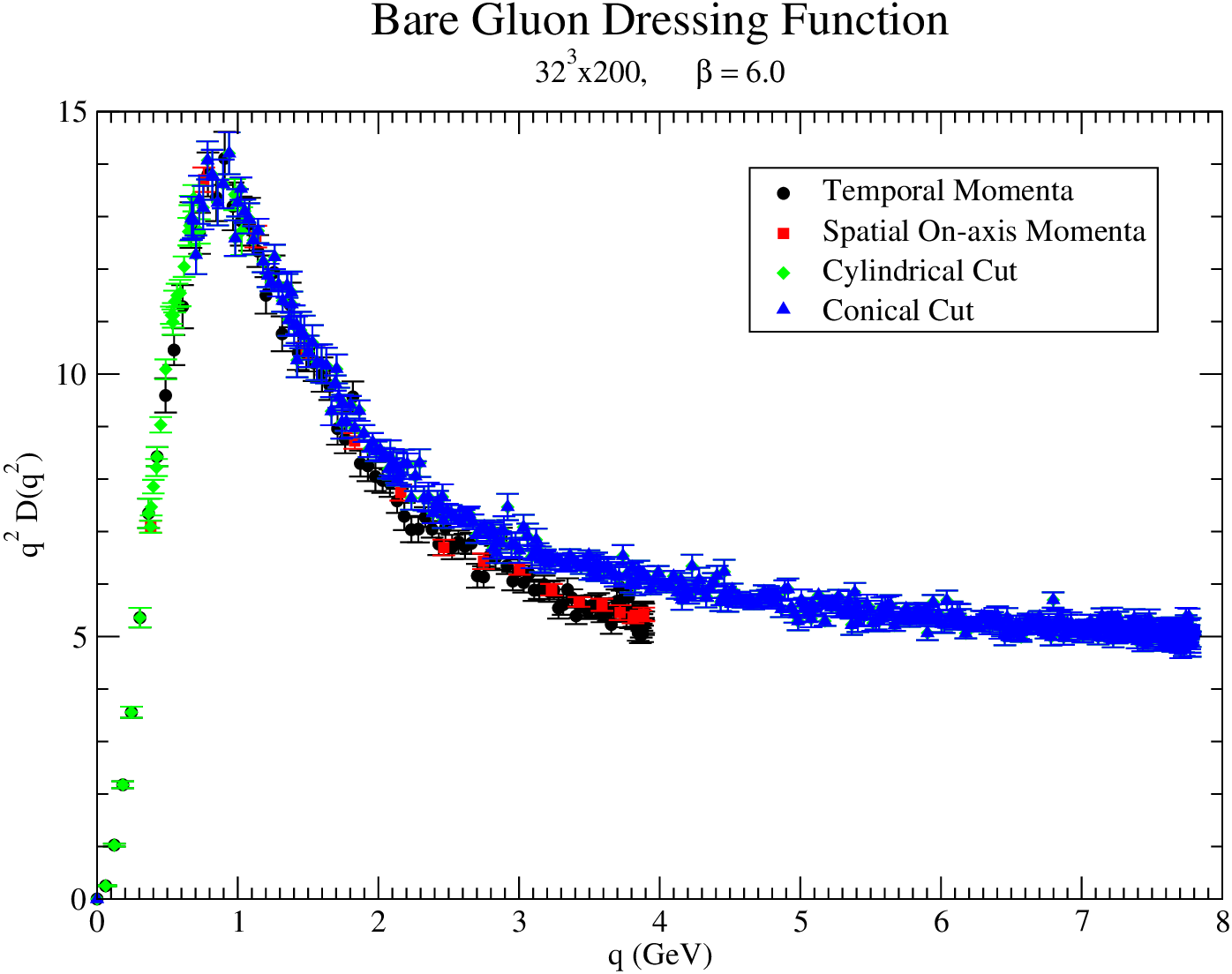}
\hfill
\includegraphics[scale=0.3]{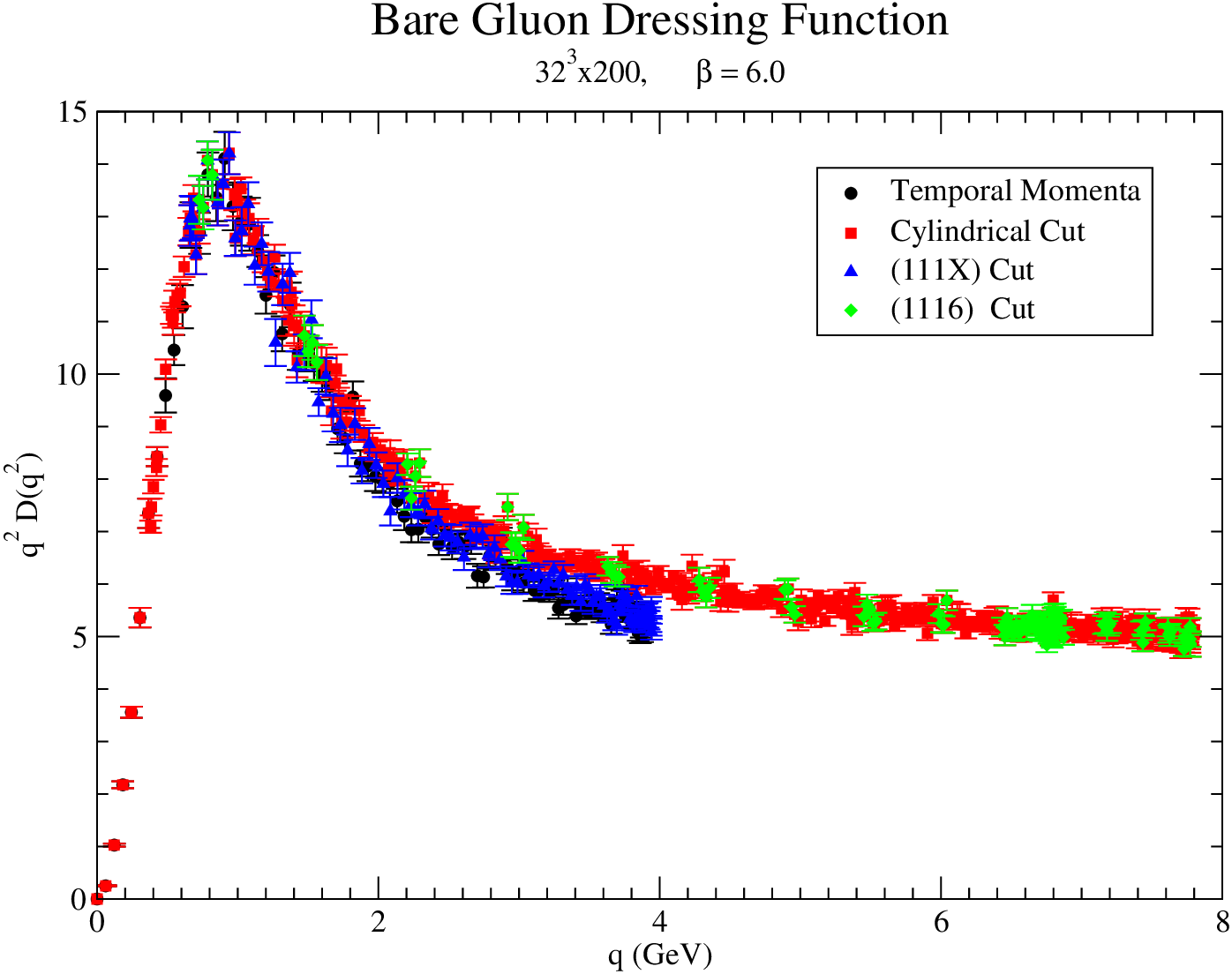}
\centering
\caption{The gluon dressing function on a $32^3 \times 200$ lattice. 
The different momentum cuts are discussed in the text.}
\label{eq:assimetrica}
\end{figure*}

The gluon propagator and dressing function were computed for 39
configurations using a $32^3 \times 200$ lattice. The data shows
discretization effects similar to those seen in the symmetric
lattices.  Indeed, it is well-known that for momenta above $q > 1$ GeV
the propagator is not a simple function of $q^2=q_{\mu}^2$ alone.  The
traditional approach is to apply momenta cuts \cite{adelaide} which
reduce the dependence of the propagator and the dressing function on
other $Z_4$ (in our case $Z_3$) invariants to a unique curve.  This is
better seen in the gluon dressing function. In order to illustrate
this effect, in Fig.~\ref{eq:assimetrica} we plot the gluon dressing
function for
different choices of momenta (purely time-like and different cuts of spatial
momenta).

The plot shows that, within our limited statistics, there is very good 
agreement between the dressing functions computed using purely spatial 
on-axis and purely temporal momenta. The figure does not include the dressing
function for all purely spatial momenta. However, 
in what concerns the purely spatial momenta, the gluon dressing function for
on-axis momenta 
evolves typically
along the lower edge of the spatial (including off-axis) momentum data.
The diagonal choice of momenta, {\it i.e.} the cylindrical cut \cite{adelaide}
where $n_{\mu} \approx \pm n_{\nu}$ (see the left plot in
Fig.~\ref{eq:assimetrica}) picks up an unique propagator which is
slightly  
above the propagator for the on-axis choice for momenta 
at $q$ larger than $\sim 1$ GeV. Note that this ``democratic'' choice
of momenta has been successfully used to suppress discretization effects 
such that data from different volumes and lattice spacings match at
larger momenta.

In the right plot of Fig.~\ref{eq:assimetrica} we also show two naive
generalizations to the asymmetric case of
the on-axis case and of
a ``democratic'' choice of momenta (the cylinder cut), 
labelled as $(111X)$ and $(1116)$ cuts. 
The former includes momenta only of type $n_\mu = (\pm1, \pm1, \pm1,
\pm n_t)$ with $n_t = 0, 1, \dots, T/2$ (softening of on-axis momenta), 
while 
the latter includes momenta defined around the direction 
$n_\mu = (\pm n, \pm n, \pm n, \pm 6n_t)$ with
$n = 0, 1, \dots, L-1$ and $n_t = n, n \pm 1, n \pm 2$. This direction is 
close to the diagonal in the elongated volume 
(remember that $T/L = 6.25$). Note that the $(111X)$ cut reproduces the 
results of the symmetric lattice for the on-axis choice of momenta, while the
$(1116)$ cut follows the symmetric lattice data for the cylindrical
and conical cuts.  
Given that the cylindrical and conical cuts seem to reduce the
finite-volume effects for momenta above 1 GeV,
Fig.~\ref{eq:assimetrica} suggests that for the asymmetric lattices
one should use the $(1116)$ cut, or a variant of it, for that
momentum region.

\section{Symmetric lattice: $32^4$}

\begin{figure}
\includegraphics[scale=0.3]{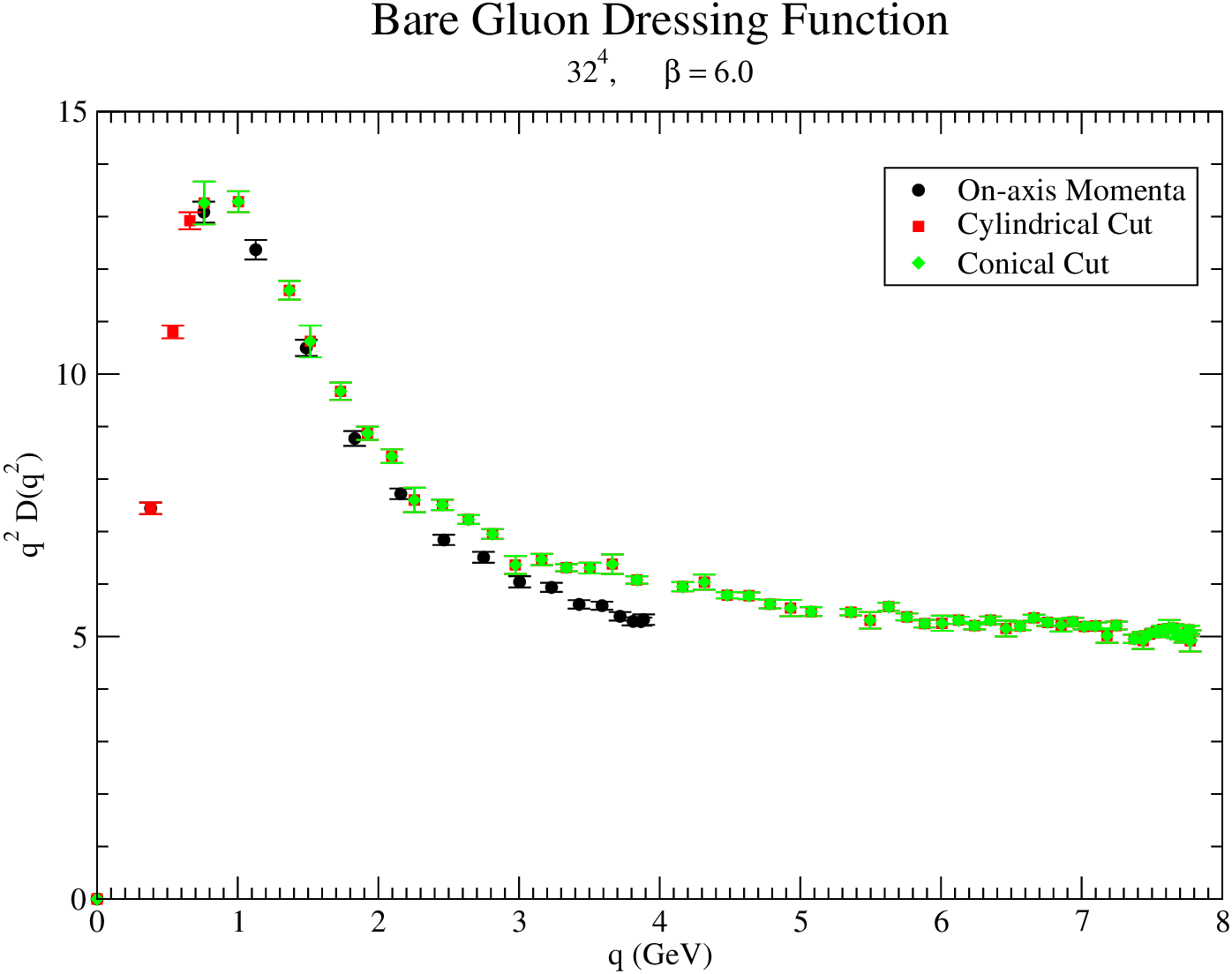}
\hfill
\includegraphics[scale=0.3]{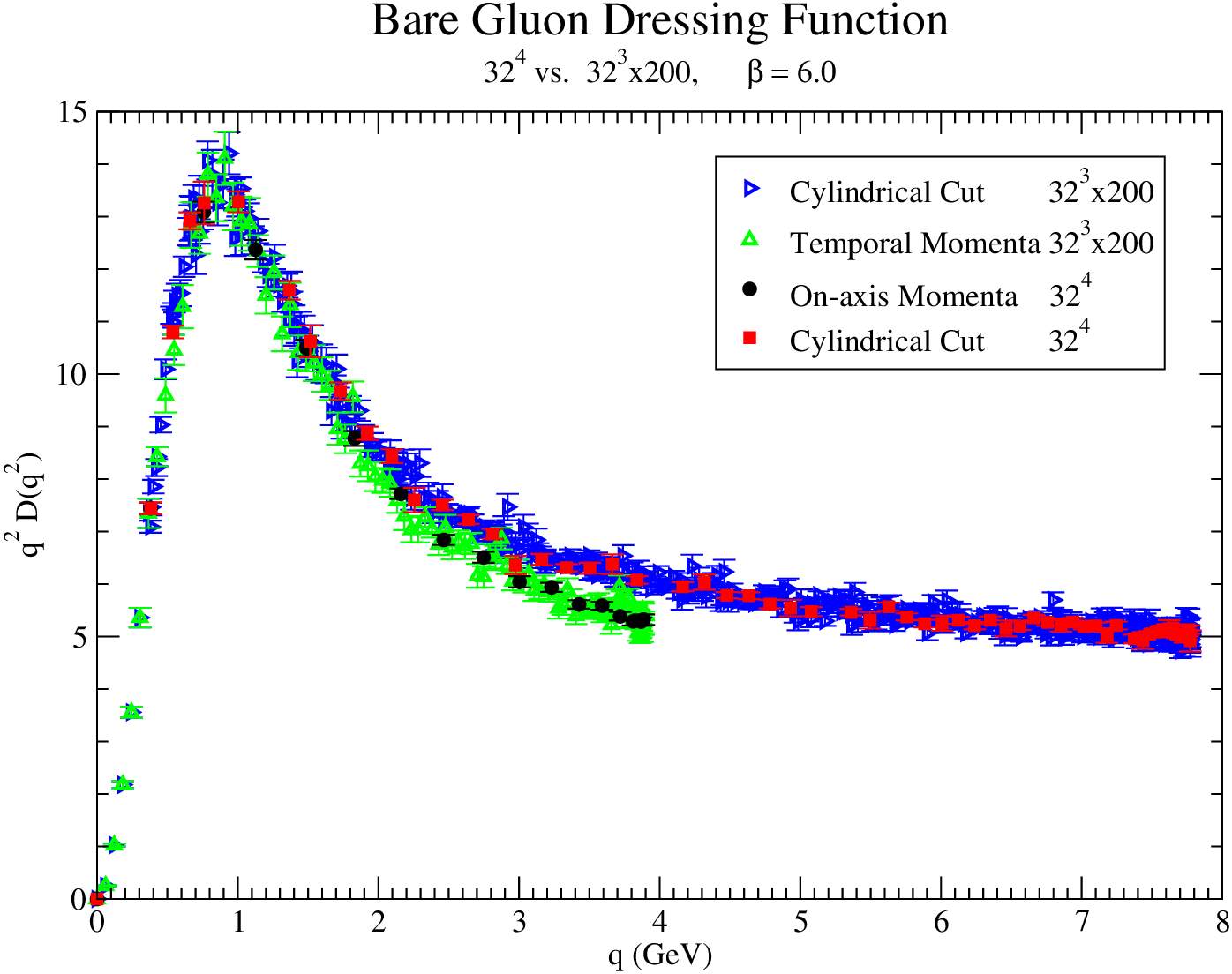}
\caption{The gluon dressing function for a $32^4$ lattice is shown on
the left hand side. The right figure compares the same data with those
on a $32^3 \times 200$ lattice. Different cuts have been applied
to the data.}
\label{eq:simetrica}
\end{figure}

The gluon dressing function $Z(q^2) = q^2 D(q^2)$, computed for an
ensemble of 50 gauge configurations on a $32^4$ lattice, is given in
the left plot of Fig.~\ref{eq:simetrica} applying various cuts. 
The lattice data shows similar discretization effects as for the
asymmetric lattice. This is illustrated in the right plot of
the same figure. 
There, data for the two lattices $32^4$ and $32^3 \times 200$ are
shown for two different momentum cuts at larger momenta and good
agreement is found. Moreover, within our limited statistics, the  
cuts produce similar effects for both lattices
where the on-axis data lie in both cases systematically below
cylinder-cut data for $q > 2$ GeV.

\begin{figure}
\centering
\includegraphics[scale=0.4]{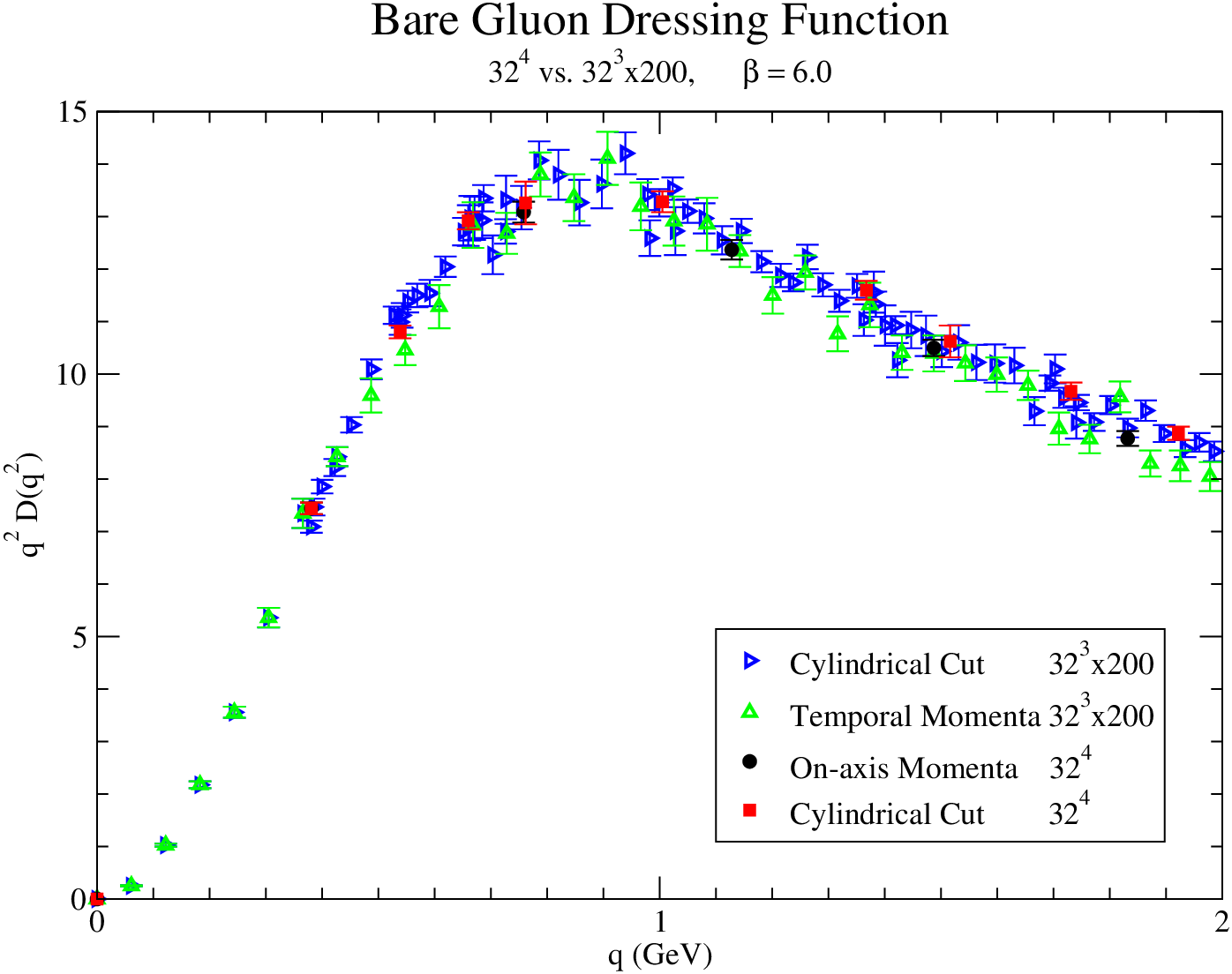}
\caption{The gluon dressing function at low momenta for the $32^4$ and
the $32^3 \times 200$ lattices for two different momenta cuts.
\label{eq:duas}}
\end{figure}

In Fig.~\ref{eq:duas} the lattice gluon dressing function for a $32^3
\times 200$ and a $32^4$ lattice are compared for momenta below 2
GeV. Again, two different cuts (on-axis and cylindrical) are
considered. We find that in this momentum range the dressing functions
$Z(q^2)$ for the two lattice geometries are in good agreement, even
though we cannot compare at the low-lying momenta. A comparison with
data from larger symmetric lattices is necessary to become more
confident in this. At least, the good matching between 0.5~GeV and 2
GeV is encouraging in what concerns the use of asymmetric lattices to
extract reliable infrared properties in future lattice simulations.

\section{The impact of the spatial volume}

\begin{figure}
\includegraphics[scale=0.3]{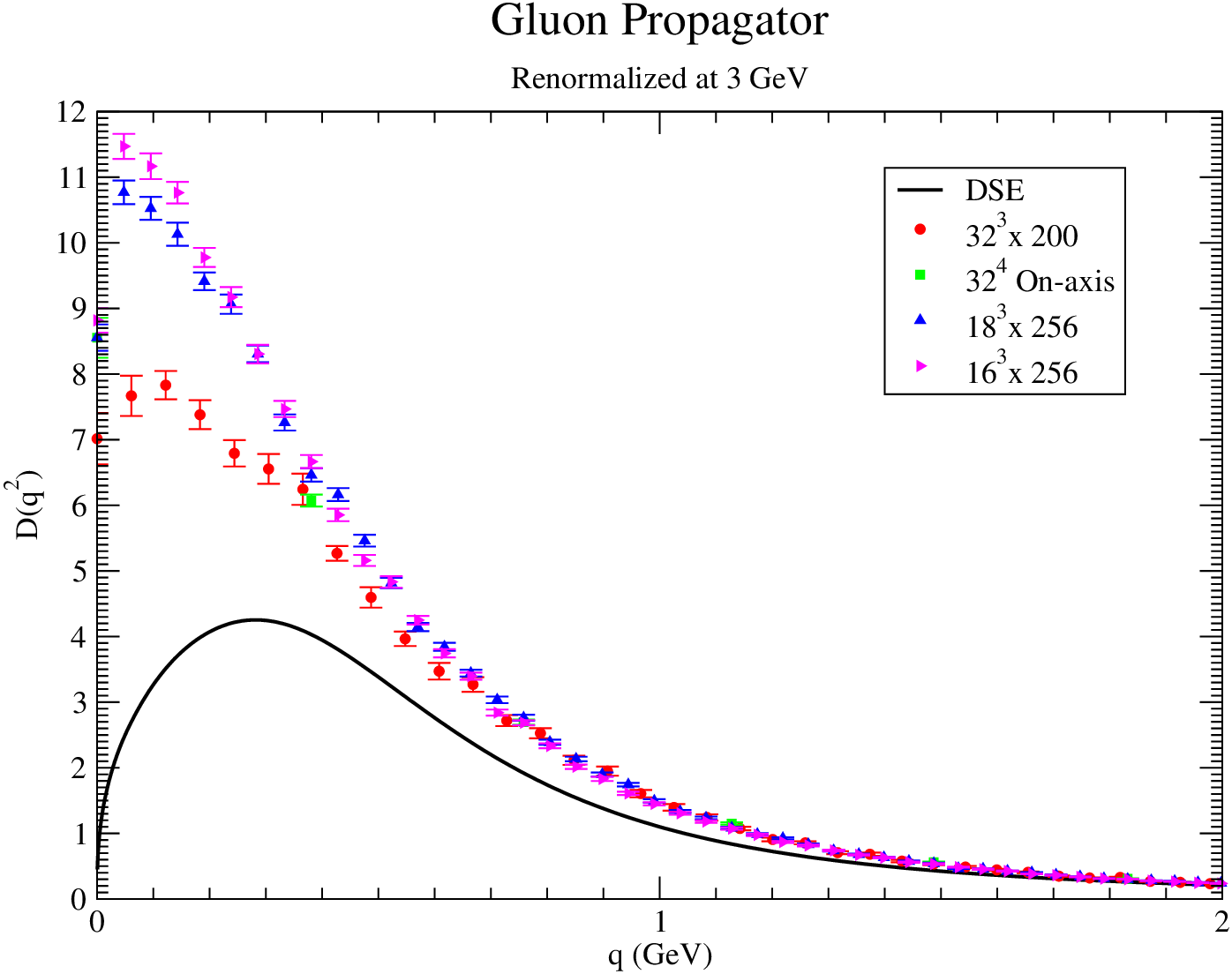}
\hfill
\includegraphics[scale=0.3]{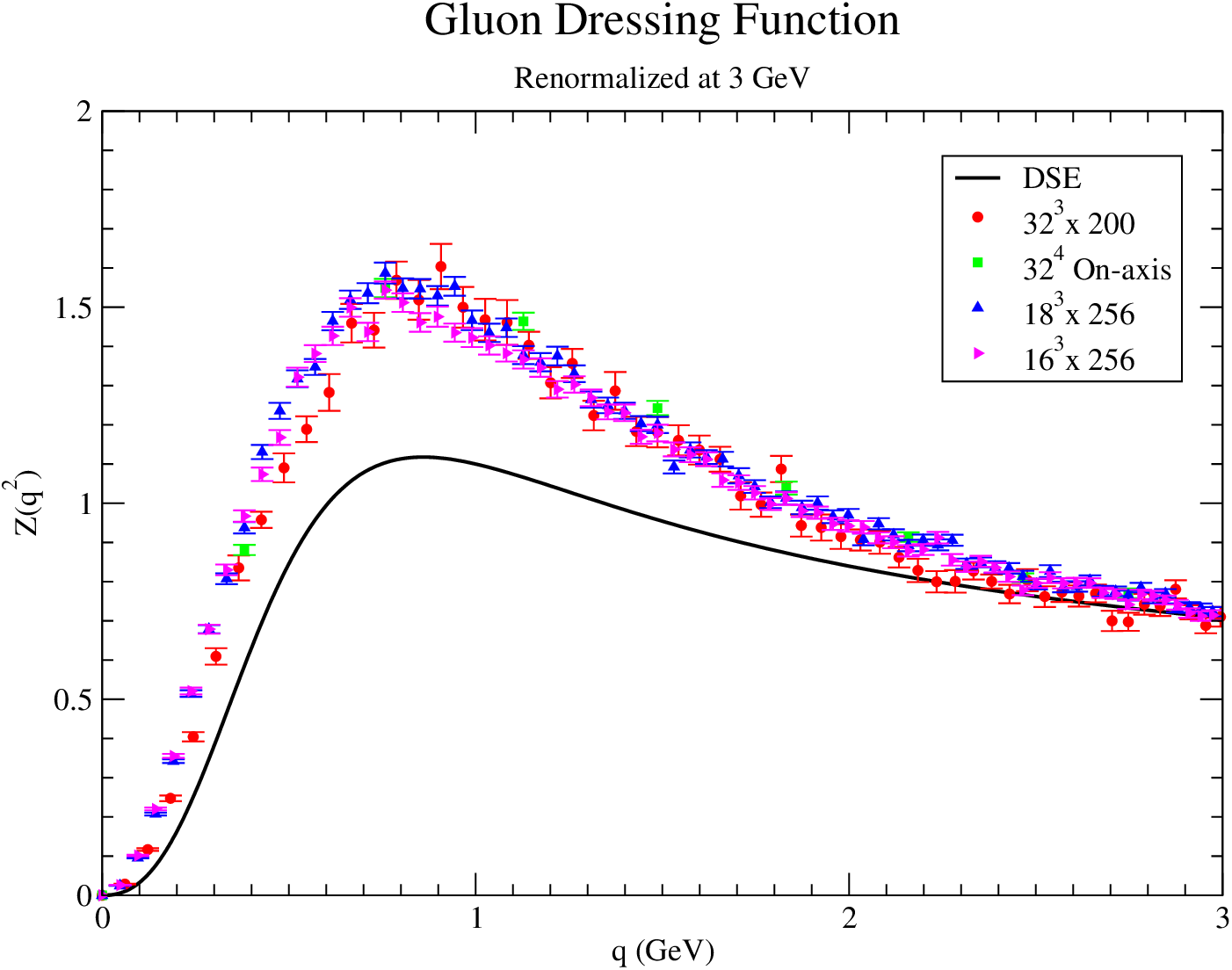}
\caption{The gluon propagator (left) and the gluon dressing function (right)
for various asymmetric lattices
and for $32^4$ at $\beta = 6.0$ . 
compared with the DSE solution.
\label{eq:varias}}
\end{figure}

Now we discuss the volume dependence of the propagator $D(q \ne 0)$
as function of the physical momentum $q$.
In Fig.~\ref{eq:varias} we compare the gluon propagator on one hand and the 
dressing function on the other hand for various 
asymmetric lattices, for $32^4$ (all at $\beta = 6.0$) and for
the continuum Dyson-Schwinger solution of Ref.~\cite{Fischer}. 
All propagators were renormalized according to the condition
\begin{equation}
   \left. D(q^2) \right|_{q^2 = \mu^2} = \frac{1}{\mu^2}, 
\end{equation}
with the choice $\mu = 3 {\rm~GeV}$. For asymmetric lattices, the data is for
time-like (on-axis) momenta. For the symmetric lattice, the plot includes
only on-axis momenta. Fig.~\ref{eq:varias} shows that the two results
become closer as the lattice volume increases.

\section{Fitting the IR gluon dressing function}

In previous investigations it was verified that the lattice gluon dressing 
function and the continuum Dyson-Schwinger solution are well described 
not by a pure power law but by both functions
\begin{equation}
   Z_I(q^2) ~ = ~ \omega \left( \frac{q^2}{q^2 + \Lambda^2} \right)^{2 \kappa}
  \, ,   
  \hspace{0.7cm}
   Z_{II} (q^2) ~ = ~ \omega \frac{ \left( q^2 \right)^{2 \kappa} }
                                  { \left( q^2 \right)^{2 \kappa} + 
                                    \left( \Lambda^2 \right)^{2 \kappa} }
\end{equation}
for momenta below $ \sim 700$ MeV. 

\begin{table}
\begin{center}
\begin{tabular}
   {r@{\hspace{0.7cm}}l@{\hspace{0.7cm}}l@{\hspace{0.7cm}}l@{\hspace{0.7cm}}ll}
\hline
   Lattice &  $q_{max}$   &  $\kappa$  & $\Lambda$  & $\chi^2 /d.o.f.$
    & \# conf \\
\hline
  $16^3 \times 256$ &  664       & $0.5090^{+19}_{-20}$ &
                                   $409^{+4}_{-4}$      & 
                                   0.71 & 155 \\
   &   &  &  &  &  \\
  $18^3 \times 256$ &  711       & $0.5320^{+28}_{-30}$  &
                                   $389^{+5}_{-6}$       & 
                                   1.14 & 150 \\
   &   &  &  &  &  \\
  $32^3 \times 200$ &  728       & $0.532^{+12}_{-12}$  & 
                                   $465^{+25}_{-23}$    &
                                   1.28 & 39 \\
\hline
\end{tabular}
\caption{Parameters corresponding to a fit of the lattice gluon dressing 
function according to $Z_I (q^2)$. $q_{max}$ and $\Lambda$ are given in MeV.
\label{tab:ZI}}
\end{center}
\end{table}

\begin{table}
\begin{center}
\begin{tabular}
  {r@{\hspace{0.7cm}}l@{\hspace{0.7cm}}l@{\hspace{0.7cm}}l@{\hspace{0.7cm}}ll}
\hline
   Lattice &  $q_{max}$   & $\kappa$         & $\Lambda$ & $\chi^2 /d.o.f.$ 
 & \# conf \\
\hline
  $16^3 \times 256$ &  664       & $0.5077^{+16}_{-17}$ &
                                   $409^{+4}_{-3}$      & 
                                   0.69 & 155 \\
   &   &  &  &  &  \\
  $18^3 \times 256$ &  711       & $0.5266^{+29}_{-21}$  &
                                   $391^{+3}_{-7}$      & 
                                   1.09 & 150 \\
   &   &  &  &  &  \\
  $32^3 \times 200$ &  728       & $0.528^{+10}_{-8}$ & 
                                   $464^{+22}_{-23}$    &
                                 1.16 & 39 \\
\hline
\end{tabular}
\caption{Parameters corresponding to a fit of the lattice gluon dressing 
function according to $Z_{II}(q^2)$. $q_{max}$ and $\Lambda$ are given in MeV.
\label{tab:ZII}}
\end{center}
\end{table}

The results of fitting the lattice dressing function for purely temporal 
momenta
with $Z_I$ and $Z_{II}$ are reported in tables \ref{tab:ZI} and \ref{tab:ZII}, 
respectively; $q_{max}$ is the highest momentum included 
in the fits. Note that, the exponent $\kappa$ agrees within one standard 
deviation for the two largest volumes. Furthermore, for these lattices, 
$\kappa$ agrees with the estimate of O.O. and P.J.S. \cite{Arx},
$\kappa \sim 0.53$, from using ratios of propagators to suppress the volume 
dependence. The ratio method discussed in~\cite{Arx}, if applied 
to the $32^3 \times 200$ data, estimates $\kappa = 0.565 \pm 0.040$ for the 
infrared exponent.

The results give $\kappa$ consistently above $0.5$. If this really represents 
the infrared asymptotics, it supports a vanishing 
$q \to 0$ limit of the gluon propagator $D(q \ne 0)$.
Moreover, one should keep in mind that the fits to a pure power
law provide always a $\kappa < 0.5$, with $\kappa$ increasing with the lattice
volume.
It should be noted, however, that so far lattice simulations have always
reported a finite and not vanishing gluon propagator at zero momentum
\cite{BraMM}.

\section{Results and Conclusions}

In this work the gluon propagator and dressing function has been
analysed for various asymmetric lattices and a comparison to $32^4$
data has been done. Despite the observed finite-volume effects, for
volumes as large as $32^3 \times 200$ the dressing function $Z(q^2)$
for purely temporal momenta agrees well, within the available
statistics, with the corresponding function for purely spatial on-axis momenta. 

In what concerns the $32^3 \times 200$ and $32^4$ data, the momentum
cuts produce similar results for the full range of 
momenta, {\it i.e.} the gluon propagator/dressing function for on-axis momenta 
are systematically below cone-cut or cylinder-cut data for $q > 1$ GeV.

The behaviour of the lattice infrared gluon dressing function $Z(q^2)$
is well described by the two ans\"atze $Z_I$ and $Z_{II}$ for $q < 700
MeV$. The fits to the data provide $\kappa$ values which support a  
vanishing zero momentum limit of the gluon propagator for all the lattices 
reported here. 
The measurement of the infrared exponent $\kappa$ 
for the two larger lattices suggests a value
$\kappa \sim 0.53$, in agreement with the estimate discussed in \cite{Arx}.

\section*{Acknowlegdements} 

P.J.S. acknowledges F.C.T. financial support via grant SFRH/BD/10740/2002.
This work was supported in part by F.C.T. under contracts POCI/FP/63436/2005
and POCI/FP/63923/2005. E.-M.~I. thanks the DFG for support through the 
Forschergruppe FOR 465. A.~S. is supported by the Australian Research Council.

\end{document}